\documentclass[12pt]{article}
\usepackage{graphicx}

\def \bea{\begin{eqnarray}}
\def \beq{\begin{equation}}

\def \eea{\end{eqnarray}}
\def \eeq{\end{equation}}

\def \s{\sqrt{2}}
\def \st{\sqrt{3}}
\def \sx{\sqrt{6}}

\begin{document}
\rightline{EFI 10-8}
\rightline{arXiv:1004.3225}
\rightline{April 2010}
\bigskip
\centerline{\bf DALITZ PLOT STRUCTURE IN $D^0 \to \pi^+ \pi^- \pi^0$}
\bigskip

\centerline{Bhubanjyoti Bhattacharya$^{a,}$\footnote{bhujyo@uchicago.edu},
Cheng-Wei Chiang$^{b,c,}$\footnote{chengwei@ncu.edu.tw} and
Jonathan L. Rosner$^{a,}$\footnote{rosner@hep.uchicago.edu}}
\bigskip
\centerline{$^a$ \it Enrico Fermi Institute and Department of Physics}
\centerline{\it University of Chicago, 5640 S. Ellis Avenue, Chicago, IL 60637}
\bigskip
\centerline{$^b$ \it Department of Physics and Center for Mathematics and
  Theoretical Physics}
\centerline{\it National Central University, Chungli, Taiwan 320, Republic of
  China}
\bigskip
\centerline{$^c$ \it Institute of Physics, Academia Sinica}
\centerline{\it Taipei, Taiwan 115, Republic of China}
\bigskip

\begin{quote}

The BaBar Collaboration has pointed out that $D^0 \to \pi^+ \pi^- \pi^0$
is dominated by an isospin-zero final state, leading to nearly complete
depletion of the Dalitz plot along all three diagonals.  In flavor-SU(3)
approaches to charmed particle decays to a light vector and a light
pseudoscalar particle, this behavior is seen, but does not appear to have
a fundamental origin.  Instead, it arises as a result of approximate
cancellation of higher-isospin combinations of several types of
amplitudes:  color-favored tree, color-suppressed tree, and exchange.
Interpretation in terms of a direct-channel effect would require an
exotic resonance, with spin, parity, and charge-congjugation eigenvalues
$J^{PC} = 0^{--}$.
 
\end{quote}

\leftline{PACS categories: 11.30.Hv, 12.39.St, 13.25.Ft, 14.40.Lb}

\section{INTRODUCTION}

A curious feature of the decay $D^0 \to \pi^+ \pi^- \pi^0$
has been pointed out by the BaBar Collaboration \cite{Aubert:2007,%
Gaspero:2008,Gaspero:2010}.  Although states of isospin zero, one, and two
are possible in principle, the Dalitz plot shows strong depopulation along
each of its symmetry axes, characteristic of a final state with isospin
zero \cite{Zemach:1964}.  In the present paper we show that this behavior
is expected in a flavor-SU(3) analysis based on a graphical language
\cite{Chau}.  It validates certain assumptions made in analyzing decays of
charmed mesons within that language, including factorization
\cite{Bhattacharya:2008ss,Bhattacharya:2009,Cheng:2010ry}.  However, it still
invites explanation at a deeper level.

We review the isospin decomposition of three-pion amplitudes in Section II
and the graphical description of charmed meson decays in Section III.  We then
apply results obtained in flavor-SU(3) fits to charmed meson decay rates to
the construction of isospin amplitudes in Section IV, finding dominance of
$I=0$ as measured experimentally \cite{Aubert:2007,Gaspero:2008,Gaspero:2010}.
We discuss reasons for this agreement in Section V and conclude in Section VI.

\section{ISOSPIN DECOMPOSITION}

We retrace steps noted in Refs.\ \cite{Aubert:2007} and \cite{Gaspero:2008},
whose conventions are slightly different from ours.  In accord with standard
usage for Clebsch-Gordan coefficients \cite{PDG:2008}, we define
\bea
\rho^+ & = & \frac{1}{\s} [\pi^+ \pi^0 - \pi^0 \pi^+]~,\\
\rho^0 & = & \frac{1}{\s} [\pi^+ \pi^- - \pi^- \pi^+]~,\\
\rho^- & = & \frac{1}{\s} [\pi^0 \pi^- - \pi^- \pi^0]~.
\eea
The BaBar papers mentioned above choose the opposite sign for $\rho^0$.

With our definitions the isospin amplitudes (again in accord with standard
usage \cite{PDG:2008}) are
\bea
A_0 & = & \frac{1}{\st} [\rho^+ \pi^- - \rho^0 \pi^0 + \rho^- \pi^+]~,\\
A_1 & = & \frac{1}{\s} [\rho^+ \pi^- - \rho^- \pi^+]~,\\
A_2 & = & \frac{1}{\sx} [\rho^+ \pi^- + 2 \rho^0 \pi^0 + \rho^- \pi^+]~,
\eea
where we have used $\rho^i \pi^j$ as a shorthand for ${\cal A}(D^0 \to
\rho^i \pi^j)$.

A purely isospin-zero spin-zero configuration of three pions consists
of each pair coupled to isospin one so as to couple with the third pion
to isospin zero.  Because of Bose statistics, each pion pair must then be
antisymmetric under interchange of the pions.  The matrix element must
then be of the form
\beq
{\cal A}(D^0 \to \pi^+ \pi^- \pi^0) = f(\mathbf{p}_+,\mathbf{p}_-,
\mathbf{p}_0)~,
\eeq
where $f$ is a totally antisymmetric function of its arguments and
$\mathbf{p}_i$ is the three-momentum of pion $\pi^i$ in the $D^0$
center-of-mass system (c.m.s.) \cite{Zemach:1964}.  The corresponding Dalitz
plot must have no events along each of its symmetry axes.

\begin{figure}
\includegraphics[width=0.96\textwidth]{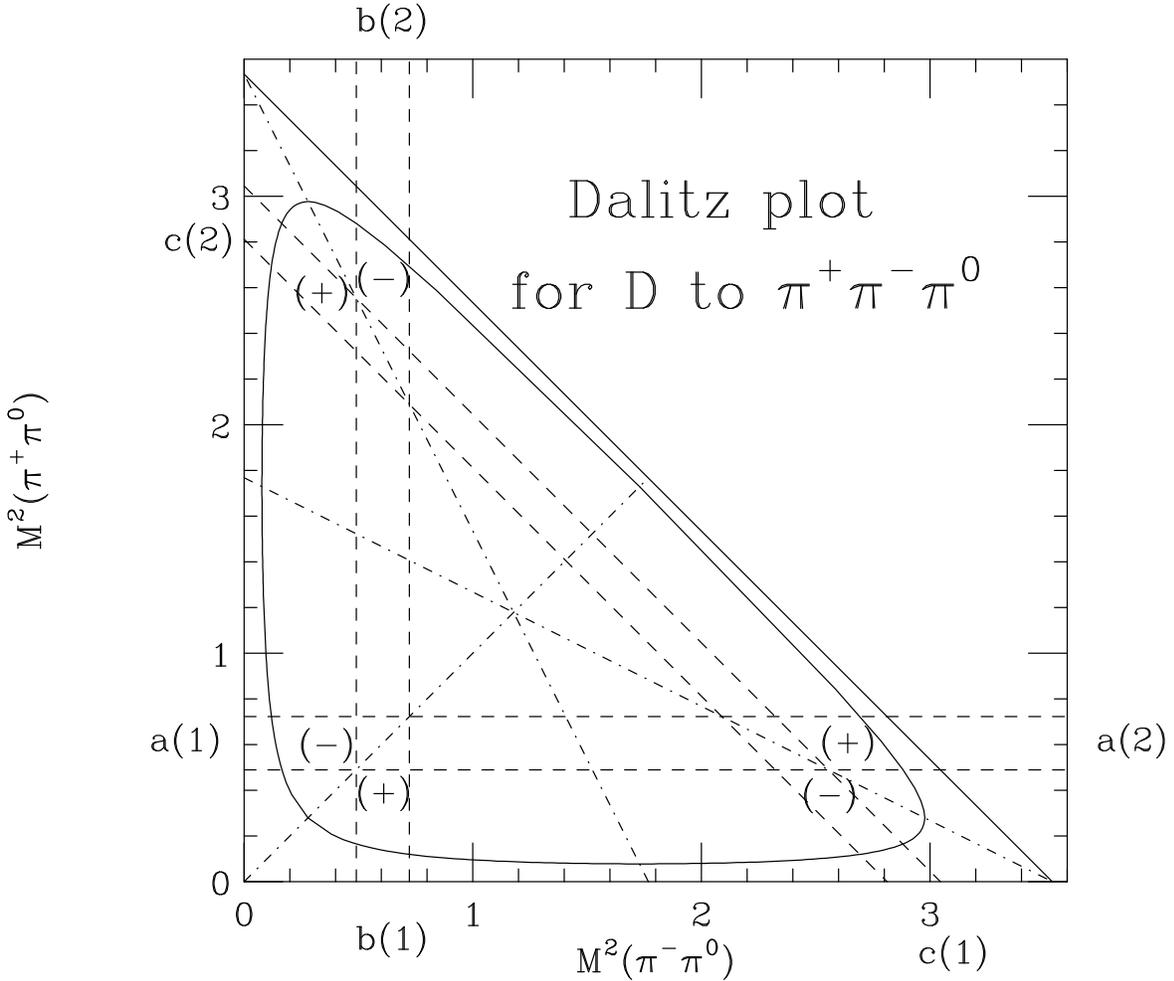}
\caption{Dalitz plot for $D^0 \to \pi^+ \pi^- \pi^0$ illustrating $\rho(770)$
bands (between pairs of dashed lines), symmetry axes (dash-dotted lines)
along which $I=0$ amplitudes vanish, and relative signs of interfering
amplitudes in regions where bands cross.  Bands are labeled by letters
(a,b,c) for ($\rho^+,\rho^-,\rho^0)$; the two ends of each band are labeled
by numbers corresponding to the configurations illustrated in Fig.\
\ref{fig:decays}.
\label{fig:dal}}
\end{figure}

\begin{figure}
\includegraphics[width=0.98\textwidth]{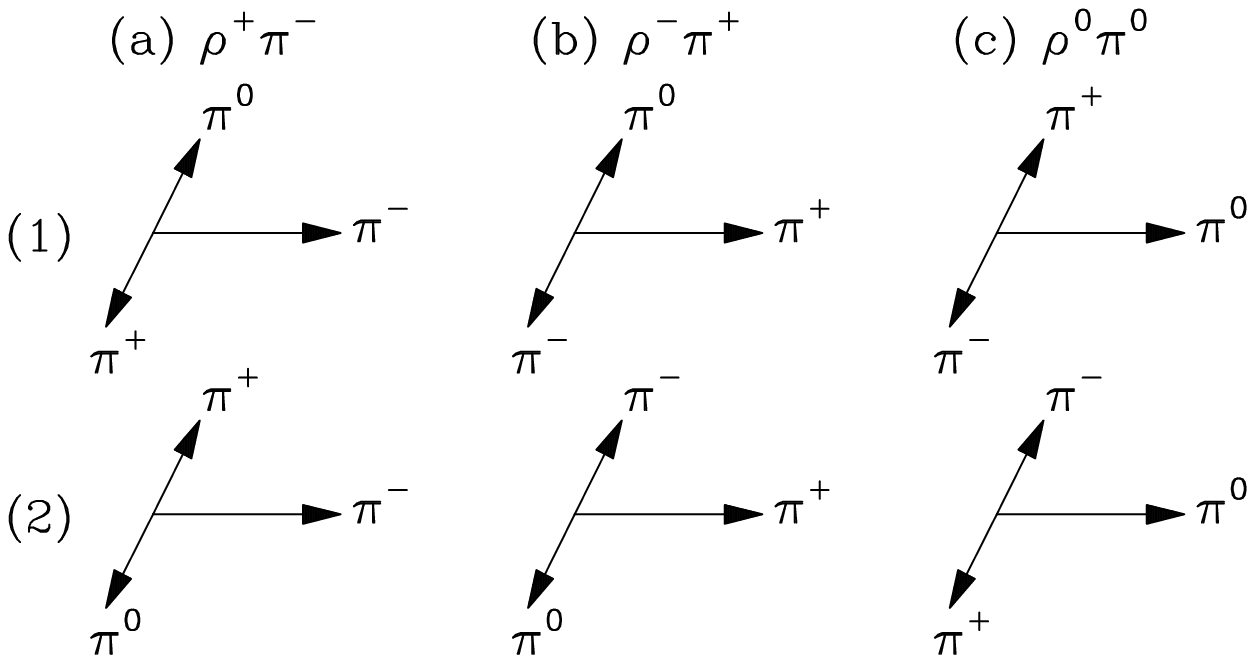}
\caption{Configurations in $D^0 \to \rho \pi$ corresponding to regions in
Fig.\ \ref{fig:dal}.  Each configuration is illustrated in the $\rho$
c.m.s.  When one of the decay products of the $\rho$ makes a small angle with
the bachelor pion, it is near the kinematic region in which the two can form
another $\rho$.
\label{fig:decays}}
\end{figure}

If $A_1 = A_2 = 0$, one must have
\beq \label{eqn:rel}
{\cal A}(D^0 \to \rho^+ \pi^-) = -{\cal A}(D^0 \to \rho^0 \pi^0) =
 {\cal A}(D^0 \to\rho^- \pi^+)~.
\eeq
Figs.\ \ref{fig:dal} and \ref{fig:decays} then permit one to see the
destructive interference along each symmetry axis of the Dalitz plot.

The (horizontal, vertical, diagonal) bands respectively denote the $(\rho^+,
\rho^-,\rho^0)$ regions.  Because each $\rho$ decays to $\pi \pi$
in a P-wave, the signs of the $\rho \pi$ amplitudes are opposite at opposite
ends of each band.  Bands and labels of each of their ends correspond to
configurations shown in Fig.\ \ref{fig:decays}.

\begin{table}
\caption{Observed amplitudes (arbitrary overall normalization) and phases for
$D^0 \to \rho(770) \pi$ \cite{Gaspero:2010}. We do not show several other
amplitudes in the fit, all of which have fractions less than a few percent. Our
convention for the phase of $A(\rho^0 \pi^0)$ differs from that of Refs.\ [1-3]
by $180^\circ$.
\label{tab:BaBar}}
\begin{center}
\begin{tabular}{c c c c} \hline \hline
Channel & Amplitude & Phase ($^\circ$) & Fraction (\%) \\ \hline
$\rho(770)^+ \pi^-$ & $0.823 \pm 0.000 \pm 0.004$ &0 (def.)& $67.8 \pm 0.0 \pm
 0.6$ \\
$\rho(770)^0 \pi^0$ & $0.512 \pm 0.005 \pm 0.011$ & $-163.8 \pm 0.6 \pm 0.4$ &
 $26.2 \pm 0.5 \pm 1.1$ \\
$\rho(770)^- \pi^+$ & $0.588 \pm 0.007 \pm 0.003$ & $-2.0 \pm 0.6 \pm 0.6$ & 
 $34.6 \pm 0.8 \pm 0.3$ \\ \hline \hline
\end{tabular}
\end{center}
\end{table}

\begin{figure}
\includegraphics[width=0.98\textwidth]{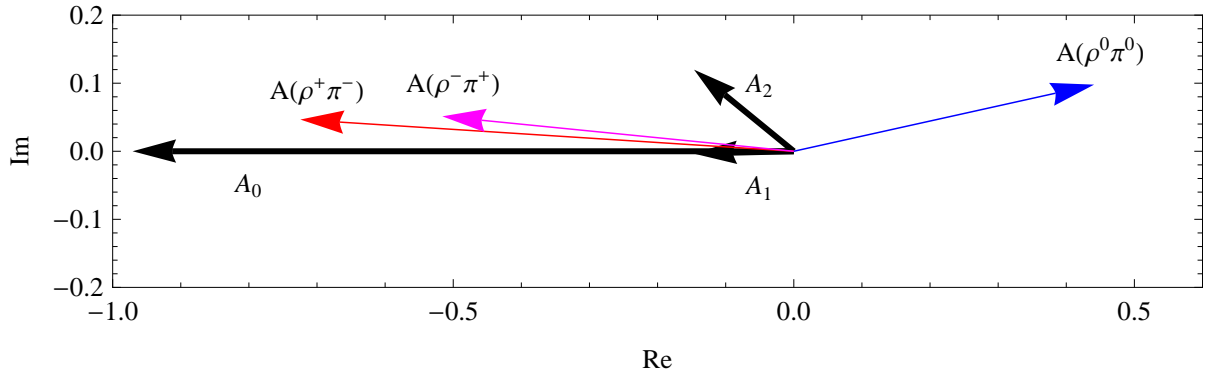}
\caption{Amplitudes and phases for $D^0 \to\rho(770) \pi$ charge states and
isospins observed in the BaBar analysis \cite{Gaspero:2010}.  Amplitudes have
been normalized so that the sums of their squares (for charge states or
isospins) is 1, and $A_0$ has been taken real and negative.
\label{fig:BaBplot}}
\end{figure}

With decay amplitudes obeying Eq.\ (\ref{eqn:rel}), and $\rho$ decays
described by Eqs.\ (1--3), each overlap region contains one amplitude
of one sign and another of the opposite sign, symptomatic of a vanishing
amplitude along each symmetry axis of the Dalitz plot.  Indeed, the
BaBar Collaboration's amplitudes shown in Table \ref{tab:BaBar}, where we have
expressed the $\rho^0 \pi^0$ amplitude in our phase convention, approximately
satisfy Eq.\ (\ref{eqn:rel}), leading to dominance of the $I=0$ amplitude as
noted in Table \ref{tab:BaBI}.  We also plot in Fig.\ \ref{fig:BaBplot} the
magnitudes and phases of the decay amplitudes for $\rho(770) \pi$ charge states
and isospins, in a phase convention where $A_0$ is real and negative.

\begin{table}
\caption{Observed isospin amplitudes and phases for $D^0 \to \rho(770) \pi$,
based on Clebsch-Gordan coefficients in Eqs.\ (4)--(6) and amplitudes quoted
in Table I.  We have normalized amplitudes so that the sum of their absolute
squares is 1.  These are not the same as the isospin amplitudes quoted in
Ref.\ \cite{Gaspero:2010}, which include contributions from higher $\rho$
states.
\label{tab:BaBI}}
\begin{center}
\begin{tabular}{c c c c} \hline \hline
Channel & Amplitude & Phase ($^\circ$) & Fraction (\%) \\ \hline
$I=0$ & $0.9708 \pm 0.0021$ & 0 (def.) & $94.24 \pm 0.40$ \\
$I=1$ & $0.1474 \pm 0.0057$ & $1.3 \pm 2.0$ & $2.17 \pm 0.17$ \\
$I=2$ & $0.1893 \pm 0.0076$ & $-39.3 \pm 13.0$ & $3.58 \pm 0.29$ \\
\hline \hline
\end{tabular}
\end{center}
\end{table}

\section{GRAPHICAL AMPLITUDE DESCRIPTION}

\begin{figure}
\mbox{\includegraphics[width=0.46\textwidth]{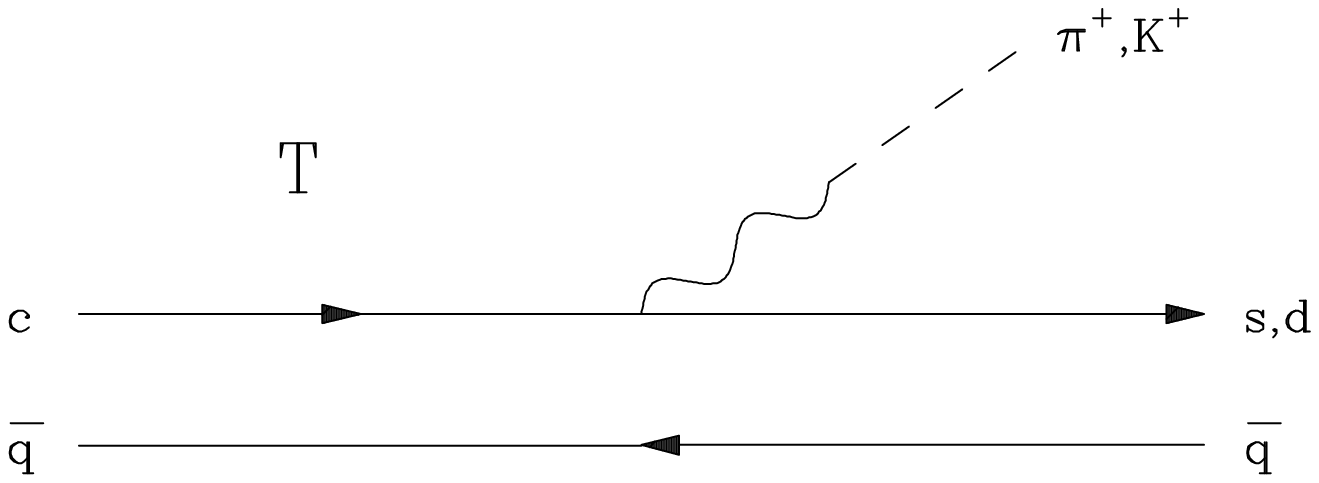} \hskip 0.3in
      \includegraphics[width=0.46\textwidth]{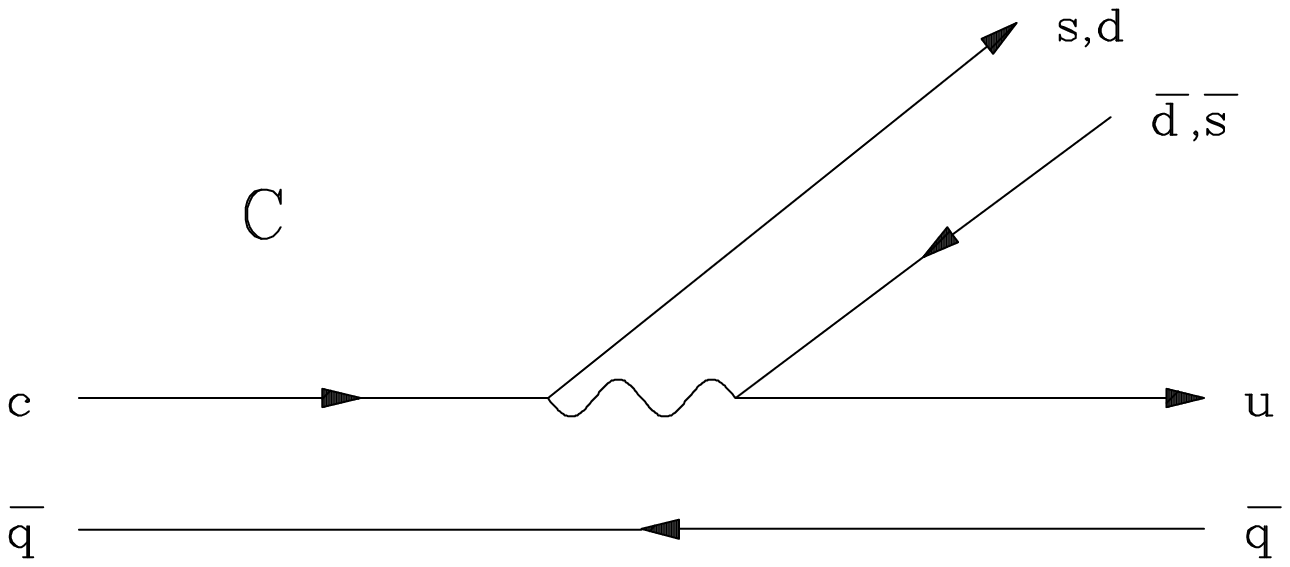}}
\vskip 0.3in
\mbox{\includegraphics[width=0.46\textwidth]{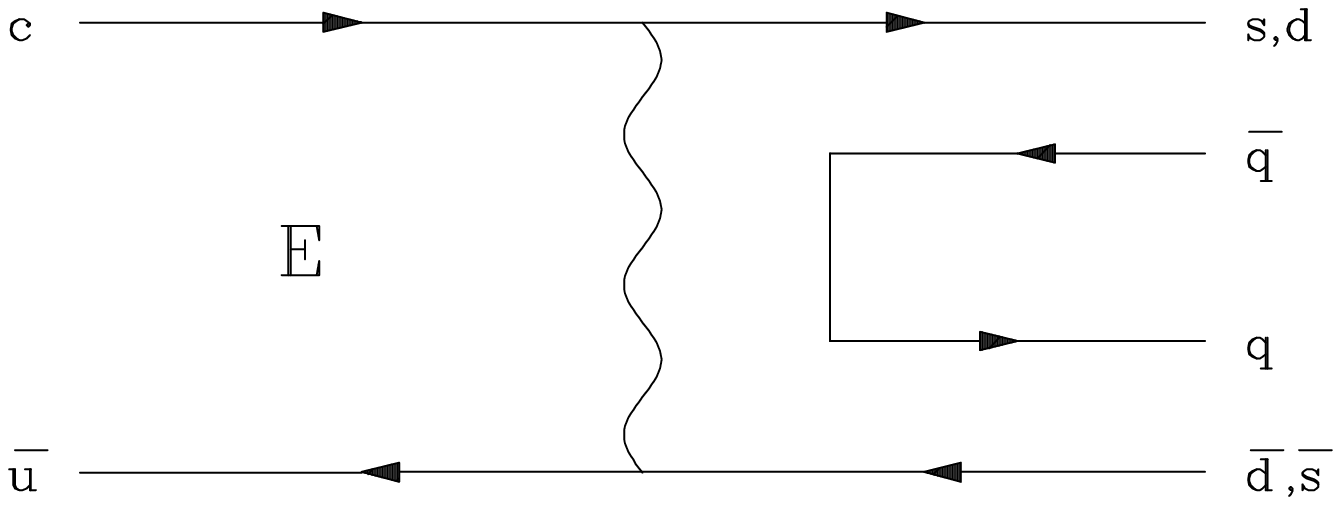} \hskip 0.3in
      \includegraphics[width=0.46\textwidth]{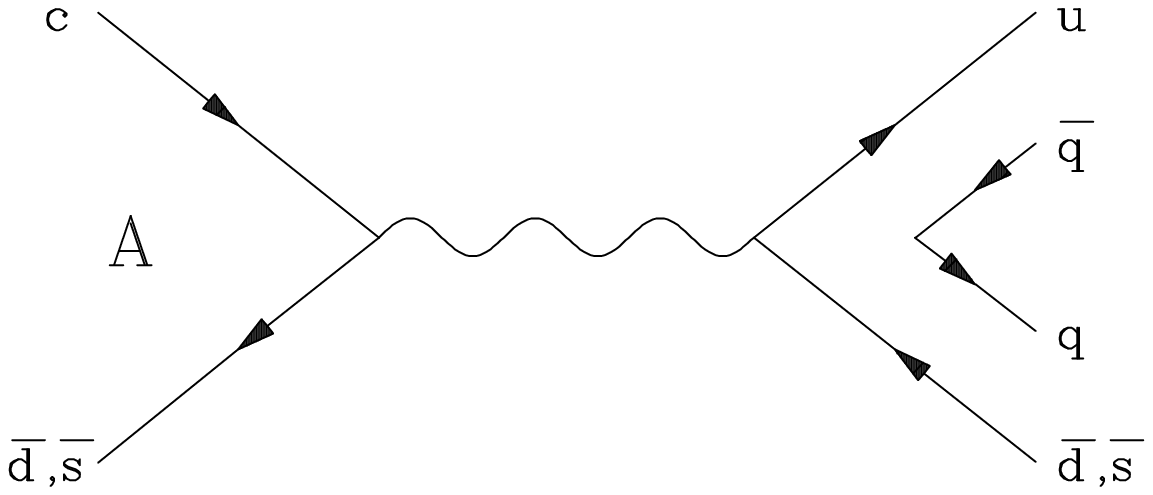}}

\caption{Flavor topologies for describing charm decays.  $T$: color-favored
tree; $C$: color-suppressed tree; $E$ exchange; $A$: annihilation.  The $D^0
\to \pi^+ \pi^- \pi^0$ decays considered here involve the CKM matrix elements
$V^*_{cd} V_{ud}$ in graphs $T$, $C$, and $E$.
\label{fig:TCEA}}
\end{figure}

Within the context of a graphical notation equivalent to flavor SU(3)
\cite{Chau}, decays of charmed mesons to pairs of light mesons
have been analyzed recently \cite{Bhattacharya:2008ss,Bhattacharya:2009,%
Cheng:2010ry}.  For decays to a pseudoscalar meson $P$ and a vector meson,
amplitudes are labeled by the type of diagram (``tree'' $T$,
``color-suppressed'' $C$, ``exchange'' $E$, and ``annihilation'' $A$), as
shown in Fig.\ \ref{fig:TCEA}.
A subscript $P$ or $V$ denotes the meson containing the spectator quark,
and a prime denotes a Cabibbo-suppressed amplitude, defined here as related
to the corresponding Cabibbo-favored amplitude by $\tan \theta_C = 0.2305$,
where $\theta_C$ is the Cabibbo angle.  (Ref.\ \cite{BM} quotes $\sin \theta_C
= 0.2246 \pm 0.0012$.)  The partial width
is given in terms of the invariant amplitude ${\cal A}$ as
\beq
\Gamma(D \to PV) = \frac{p^{*3}}{8 \pi M_D^2}|{\cal A}|^2~,
\eeq
where $p^*$ is the magnitude of the c.m.s. 3-momentum of each final particle.
The relevant decay amplitudes are then
\bea
{\cal A}(D^0 \to \rho^+ \pi^-) & = & -(T'_P + E'_V)~,\\
{\cal A}(D^0 \to \rho^- \pi^+) & = & -(T'_V + E'_P)~,\\
{\cal A}(D^0 \to \rho^0 \pi^0) & = & \frac{1}{2}(E'_P+E'_V-C'_P-C'_V)~.
\eea
The magnitudes and phases of these amplitudes obtained in flavor-SU(3) fits
to a large number of Cabibbo-favored $D^0$, $D^+$, and $D_s$ decays to $PV$
final states are summarized in Table \ref{tab:fitamps}.  Two different sets of
values are quoted, corresponding to Refs.\ \cite{Bhattacharya:2009} and
\cite{Cheng:2010ry}.  They differ mainly in their handling of $\eta$--$\eta'$
mixing, leading to different values of $E'_V$, and to a lesser extent $T'_P$
and $C'_V$.  Each amplitude is quoted in units of $10^{-6}$.

\begin{table}
\caption{Amplitudes obtained in fits to Cabibbo-favored decays of charmed
particles to $PV$ final states, in units of $10^{-6}$.
\label{tab:fitamps}}
\begin{center}
\begin{tabular}{c c c c c} \hline \hline
Amplitude & \multicolumn{2}{c}{Ref.\ \cite{Bhattacharya:2009}} &
 \multicolumn{2}{c}{Ref.\ \cite{Cheng:2010ry}} \\
       & Magnitude & Phase ($^\circ$) & Magnitude & Phase ($^\circ$) \\ \hline
$T'_P$ & $1.719\pm0.048$ & 0 (def.) & $1.776\pm0.083$ & 0 (def.) \\
$T'_V$ & $0.910\pm0.016$ & 0 (def.) & $0.910\pm0.036$ & 0 (def.) \\
$C'_V$ & $0.797\pm0.041$ & $172\pm3$ & $0.909\pm0.100$ & $164\pm23$ \\
$C'_P$ & $1.125\pm0.035$ & $-162\pm1$ & $1.126\pm0.070$ & $-162\pm3$ \\
$E'_V$ & $0.546\pm0.044$ & $-110\pm4$ & $0.330\pm0.183$ & $-124\pm41$ \\
$E'_P$ & $0.678\pm0.021$ & $-93\pm3$ & $0.677\pm0.023$ & $-93\pm5$ \\
\hline \hline
\end{tabular}
\end{center}
\end{table}

\section{PREDICTED ISOSPIN AMPLITUDES}

\begin{figure}
\includegraphics[width=0.98\textwidth]{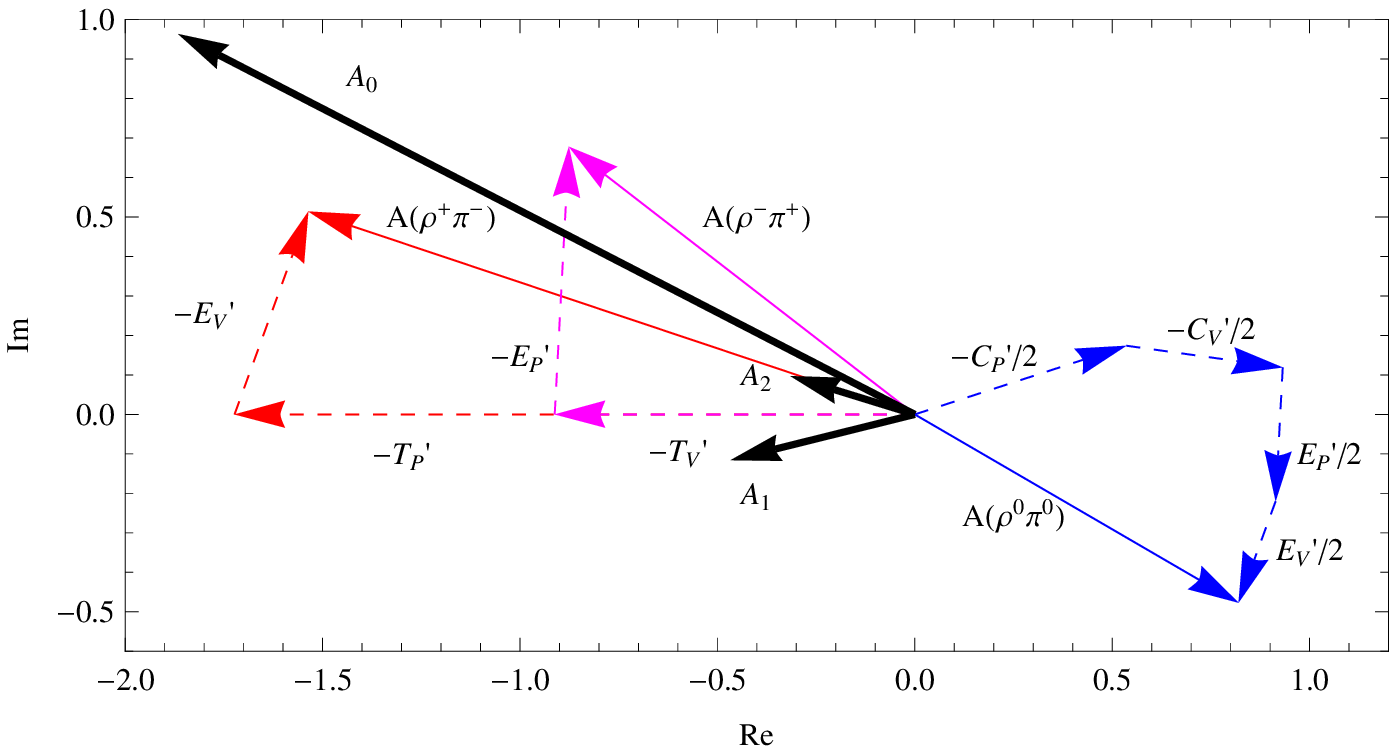}
\includegraphics[width=0.98\textwidth]{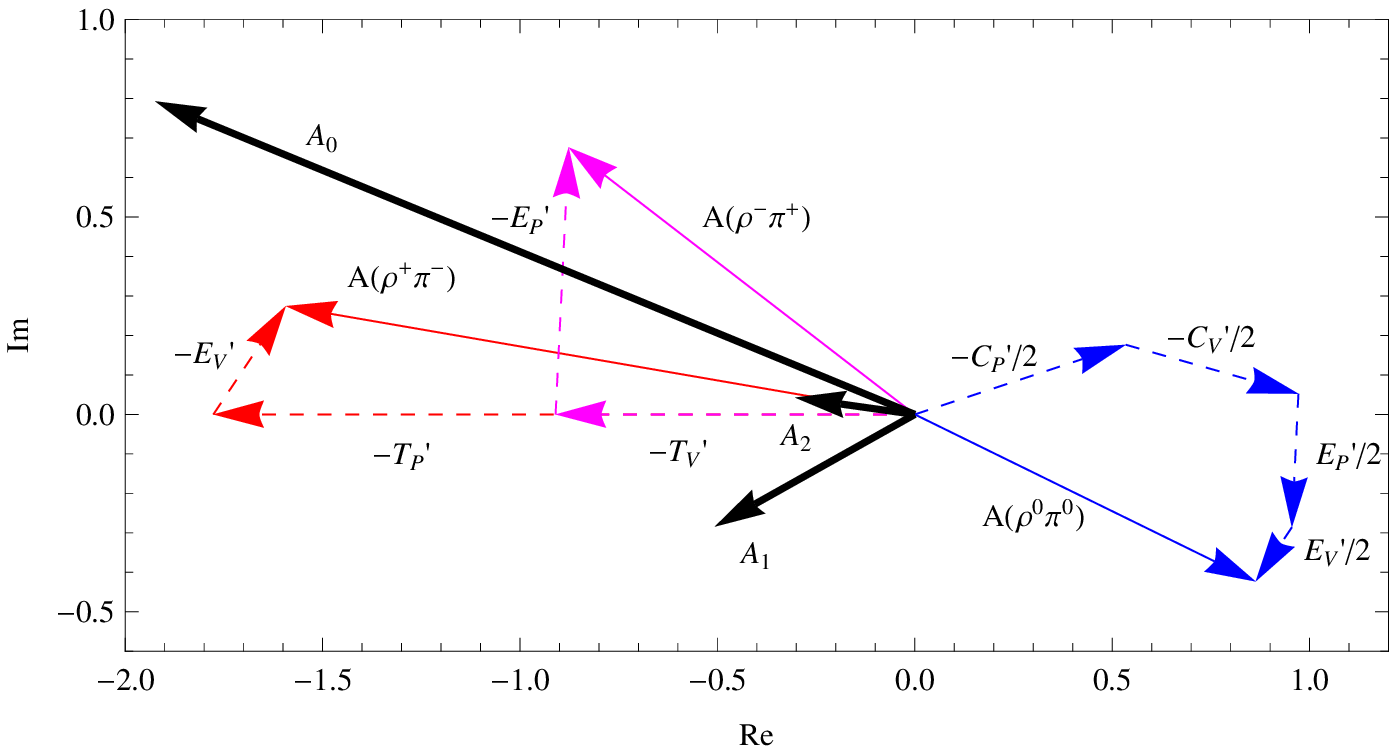}
\caption{Summing graphical amplitudes to obtain predictions for $D^0 \to \rho
\pi$ and corresponding isospin amplitudes.  Top:  Ref.\
\cite{Bhattacharya:2009}; bottom: Ref.\ \cite{Cheng:2010ry}
\label{fig:addup}}
\end{figure}

The amplitudes in Table \ref{tab:fitamps} may now be added up to give the
contributions to the different $D^0 \to \rho \pi$ decays.  The results are
shown in Fig.\ \ref{fig:addup} for the fits in Refs.\ \cite{Bhattacharya:2009}
and \cite{Cheng:2010ry}.  Also shown are the corresponding isospin amplitudes.
We give in Tables \ref{tab:BRrhopi}, \ref{tab:BRiso}, \ref{tab:CCrhopi}, and
\ref{tab:CCiso} the corresponding numerical values and their estimated errors.
One obtains dominance of isospin zero, though not quite as fully as in the
BaBar data.  The difference appears to be mainly in the $I=1$ amplitude, whose
suppression is not predicted to be as extreme as is observed.  It is in this
amplitude that the greatest difference is seen between the fits of Refs.\
\cite{Bhattacharya:2009} and \cite{Cheng:2010ry}.

\begin{table}[h]
\caption{Predicted amplitudes (in units of $10^{-6}$) and phases for $D^0 \to
\rho \pi$ in the fit of Ref.\ \cite{Bhattacharya:2009}.  Fit fractions are
defined to sum to 100\%, and compared to those in Table \ref{tab:BaBar}
normalized to 100\% in the last column.
\label{tab:BRrhopi}}
\begin{center}
\begin{tabular}{c c c c c} \hline \hline
Channel & Amplitude & Phase ($^\circ$) & Fraction (\%) & vs.\ BaBar (\%) \\
 \hline
$\rho^+ \pi^-$ & $1.615 \pm 0.058$ & $161 \pm 2$ & $55.2\pm4.0$&$52.7\pm 0.5$\\
$\rho^0 \pi^0$ & $0.946\pm0.036$ & $-30 \pm 2$& $18.9\pm 1.4$ & $20.4\pm 0.9$\\
$\rho^- \pi^+$ & $1.107\pm0.038$ & $142 \pm 2$ & $25.9\pm1.8$ & $26.9\pm 0.7$\\
 \hline \hline
\end{tabular}
\end{center}
\end{table}

\begin{table}[h]
\caption{Predicted isospin amplitudes (in units of $10^{-6}$) and phases for
$D^0 \to \rho \pi$ in the fit of Ref.\ \cite{Bhattacharya:2009}.  Fit fractions
are defined as in Table \ref{tab:BRrhopi}.
\label{tab:BRiso}}
\begin{center}
\begin{tabular}{c c c c c} \hline \hline
Channel & Amplitude & Phase ($^\circ$) & Fraction (\%) & vs.\ BaBar (\%) \\
 \hline
$I=0$ & $2.097 \pm 0.076$ & $153 \pm 1$ & $92.9\pm 6.7$& $94.24 \pm 0.40$ \\
$I=1$ & $0.477 \pm 0.015$ & $-166 \pm 2$ & $4.8 \pm 0.3$ & $2.17 \pm 0.17$ \\
$I=2$ & $0.330 \pm 0.057$ & $162 \pm 5$ & $2.3 \pm 0.8$ & $3.58 \pm 0.29$ \\
\hline \hline
\end{tabular}
\end{center}
\end{table}

\begin{table}[h]
\caption{Predicted amplitudes (in units of $10^{-6}$) and phases for $D^0 \to
\rho \pi$ in the fit of Ref.\ \cite{Cheng:2010ry}.  Fit fractions are defined
as in Table \ref{tab:BRrhopi}.
\label{tab:CCrhopi}}
\begin{center}
\begin{tabular}{c c c c c} \hline \hline
Channel & Amplitude & Phase ($^\circ$) & Fraction (\%) & vs.\ BaBar (\%) \\
 \hline
$\rho^+ \pi^-$ & $1.62\pm0.23$ & $170\pm7$ & $54.9\pm15.9$ & $52.7 \pm 0.5$\\
$\rho^0 \pi^0$ & $0.96\pm0.15$ & $-26\pm12$& $19.4\pm6.2$ & $20.4 \pm 0.9$\\
$\rho^- \pi^+$ & $1.11\pm0.06$ & $142\pm 2$& $25.7\pm2.6$ & $26.9 \pm 0.7$\\
 \hline \hline
\end{tabular}
\end{center}
\end{table}

\begin{table}[h]
\caption{Predicted isospin amplitudes (in units of $10^{-6}$) and phases for
$D^0 \to \rho \pi$ in the fit of Ref.\ \cite{Cheng:2010ry}.  Fit fractions are
defined as in Table \ref{tab:BRrhopi}.
\label{tab:CCiso}}
\begin{center}
\begin{tabular}{c c c c c} \hline \hline
Channel & Amplitude & Phase ($^\circ$) & Fraction (\%) & vs.\ BaBar (\%) \\
 \hline
$I=0$ & $2.09 \pm 0.21$ & $158 \pm 6$ & $90.9\pm18.2$ & $94.24 \pm 0.40$ \\
$I=1$ & $0.58 \pm 0.17$ & $-151\pm15$ & $7.1 \pm 4.1$ & $2.17 \pm 0.17$ \\
$I=2$ & $0.31 \pm 0.08$ & $172 \pm27$ & $2.0 \pm 1.0$ & $3.58 \pm 0.29$ \\
\hline \hline
\end{tabular}
\end{center}
\end{table}

\section{FEATURES OF $I=0$ DOMINANCE}

In the flavor-SU(3) approach one can note several regularities which contribute
to the suppression of $I=1$ and $I=2$ amplitudes.

(1) The tree amplitudes were assumed in Refs.\ \cite{Bhattacharya:2009} and
\cite{Cheng:2010ry} to be real and positive, in accord with the expectation
from factorization.  To the extent that they are equal, their contributions
cancel in the $I=1$ amplitude.  Equality would imply relations between form
factors and coupling constants; one understands $|T'_P| > |T'_V|$ on the
basis of $f_\rho > f_\pi$.

(2) The exchange amplitudes cannot contribute to $I=2$ and thus their
contributions must (and do) cancel exactly.

Some remaining effects seem more accidental:

(3) The exchange amplitudes, having the same phases, add destructively in
the $I=1$ amplitude and constructively in the $I=0$ amplitude.

(4) The color-suppressed amplitudes approximately cancel the tree amplitudes
in the $I=2$ channel.  (They do not contribute at all to $I=1$ as they
only contribute to $\rho^0 \pi^0$.)

One might be tempted to blame the enhancement of $I=0$ on a direct-channel
resonance (which would then force the relative phases of various
flavor-SU(3) amplitudes to conspire to enhance the $I=0$ channel).  However,
a three-pion final state has odd $G$-parity.  If it has $I=0$ it also must
have odd $C$-parity.  But a state with spin-parity-charge-conjugation
eigenvalue $J^{PC} = 0^{--}$ is {\it exotic}, i.e., it cannot be made of
a quark-antiquark pair.  If such a resonance exists near the $D^0$, it
must be of an unusual type, such as a hybrid or tetraquark.

\section{CONCLUSIONS}

The dominance of the isospin-zero channel in $D^0 \to \pi^+ \pi^- \pi^0$
is reproduced in SU(3)-flavor fits to charmed meson decays, with some
insight into the suppression of certain contributions to $I=1$ and $I=2$
channels.  The fact that the relative phase of the $D^0 \to \rho^+ \pi^-$
and $D^0 \to \rho^- \pi^+$ amplitudes is small supports the idea of
factorization whereby the dominant tree ($T'$) amplitudes are in phase with
one another.  This leads to a cancellation in the $I=1$ channel.  The
approximate cancellation of the $D^0 \to \rho^\pm \pi^\mp$ amplitude by
the $D^0 \to \rho^0 \pi^0$ amplitude in the $I=2$ channel requires a
cooperation between color-suppressed ($C'$) and $T'$ amplitudes.

A simple overall explanation of the remarkably simple
Dalitz plot structure (with strong depletion along each symmetry axis)
remains elusive.  It would be interesting to exhibit other Dalitz plots
with such simple structure to see if there are some common features we
might have overlooked.  The invocation of a direct-channel resonance is
tempting, but demands an as-yet-unobserved exotic state.

\section*{ACKNOWLEDGMENTS}

We thank M. Gaspero, K. Mishra, B. Meadows, and A. Soffer for helpful
communications.  This work was supported in part by the United States Department
of Energy through Grant No.\ DE FG02 90ER40560 and by the National Science
Council of Taiwan, R.~O.~C.\ under Grant No.~NSC~97-2112-M-008-002-MY3.
J.L.R. would like to thank the National Center for Theoretical Sciences for
extending hospitality.

\end{document}